\documentclass[pre,letterpaper,showpacs,showkeys,superscriptaddress,nofootinbib,twocolumn,floatfix]{revtex4}
\usepackage{amsfonts}
\usepackage{graphicx,color,amsmath,amssymb}

\newcommand{\dd}{\mathrm{d}}

\newcommand{\bvec}[1]{\ensuremath{\boldsymbol{#1}}}
\newcommand{\GeV}{\mathrm{GeV}}

\newcommand{\fm}{\mathrm{fm}}

\begin{document}

\title{Relativistic Langevin Dynamics in Expanding Media}

\author{Min He}
\affiliation{Cyclotron Institute and Department of Physics \& Astronomy,
  Texas A\&M University, College Station, TX 77843, USA}
\affiliation{Department of Applied Physics, Nanjing University of Science 
and Technology, Nanjing 210094, China}

\author{Hendrik van Hees} 
\affiliation{Frankfurt Institute for Advanced Studies,
  Ruth-Moufang-Stra{\ss}e 1, D-60438 Frankfurt, Germany}

\author{Pol B.~Gossiaux}
\affiliation{SUBATECH, UMR 6457, Laboratoire de Physique Subatomique et
  des Technologies Associ\'ees, University of Nantes - IN2P3/CNRS -
  Ecole des Mines de Nantes, 4 rue Alfred Kastler, F-44072 Nantes Cedex
  03, France}

\author{Rainer J.~Fries} 
\affiliation{Cyclotron Institute and Department of Physics \& Astronomy,
  Texas A\&M University, College Station, TX 77843, USA}

\author{Ralf Rapp}
\affiliation{Cyclotron Institute and Department of Physics \& Astronomy,
  Texas A\&M University, College Station, TX 77843, USA}

\date{\today}

\begin{abstract}
  We study the consequences of different realizations of diffusion
  processes in relativistic Langevin simulations. We elaborate on the
  Ito-Stratonovich dilemma by showing how microscopically calculated
  transport coefficients as obtained from a Boltzmann/Fokker-Planck
  equation can be implemented to lead to an unambiguous realization of
  the Langevin process. Pertinent examples within the pre-point (Ito)
  and post-point (H{\"a}nggi-Klimontovich) Langevin prescriptions are
  worked out explicitly.  Deviations from this implementation are shown
  to generate variants of the Boltzmann distribution as the stationary
  (equilibrium) solutions. Finally, we explicitly verify how the Lorentz
  invariance of the Langevin process is maintained in the presence of an
  expanding medium, including the case of an ``elliptic flow''
  transmitted to a Brownian test particle. This is particularly relevant
  for using heavy-flavor diffusion as a quantitative tool to diagnose
  transport properties of QCD matter as created in ultrarelativistic
  heavy-ion collisions.
\end{abstract}


\maketitle

\section{Introduction}
\label{sec_intro}

Since its introduction more than 100 years ago, Brownian motion has
remained a valuable and versatile tool to study a wide variety of
physical systems. In recent years considerable efforts have been devoted
to its extension and reliable implementation for relativistic
systems~\cite{GrootLeeuwenWeert1980,DunkeHanggi2005ab,Dunkel:2009tla,Debbasch1997,Kremer2007}.
In particular, it has been applied in the studies of heavy-quark (HQ)
diffusion in the Quark-Gluon Plasma
(QGP)~\cite{Svetitsky1988,vanHees:2004gq,MooreTeaney2005,Mustafa:2004dr,Gossiaux:2004qw},
in order to evaluate transport properties of the medium produced in
ultrarelativistic heavy-ion collisions (URHICs) (see, e.g.,
Ref.~\cite{Rapp:2009my} for a recent review).  Since charm- and
bottom-quark masses are much larger than the typical temperatures, as
well as the constituent masses of the equilibrated medium, a separation
of the HQ relaxation time and collision time
emerges~\cite{MooreTeaney2005,Rapp:2009my}, thus justifying a
soft-collision approximation to be accommodated by Fokker-Planck
dynamics~\cite{Landau1981}. In practice, the Fokker-Planck equation is
routinely realized by stochastic Langevin
processes~\cite{NGvanKampen2003,DunkeHanggi2005ab,Dunkel:2009tla,Rapp:2009my}.
However, it is known that the implementation of the Langevin equation is
not unique but depends on the realization of the stochastic integral,
resulting in (seemingly) different Fokker-Planck equations, known as the
``Ito-Stratonovich
dilemma''~\cite{NGvanKampen2003,DunkeHanggi2005ab,Dunkel:2009tla}.
Moreover, the ambiguities in the Langevin discretization scheme raise
concerns on the asymptotic phase-space distribution of the relativistic
Brownian particles under consideration. The latter has been an issue of
debate (see Ref.~\cite{CuberoHanggi2007} and references therein) but
should be constrained by the long-time limit of
equilibrium~\cite{Landau1981,Walton:1999dy,Arnold:2000ab,Dunkel:2009tla},
i.e., by a ``detailed-balancing'' of the drag and diffusion terms of the
underlying Fokker-Planck equation.

The large collective flow of the medium created in URHICs (reaching
collective velocities in excess of half the speed of light), and in
particular subtle angular modulations thereof (such as the elliptic
flow), make a good understanding of differences in the realization of
\emph{relativistic} Langevin simulations mandatory. In particular, the
description of the phase-space distribution of a Brownian particle (the
heavy quark), which is usually formulated in laboratory-frame
coordinates, needs to manifestly recover the equilibrium limit given by
the Lorentz-invariant Boltzmann-J{\"u}ttner
distribution~\cite{NGvanKampen1969,Debbasch2001}.  This is evident as
the underlying Fokker-Planck equation is to yield a good approximation
of the Lorentz-covariant Boltzmann
equation~\cite{Landau1981,GrootLeeuwenWeert1980,NGvanKampen2003}.  This
problem has been addressed, e.g., in
Refs.~\cite{Walton:1999dy,Dunkel:2008jk} within the so-called Ito
realization, while pertinent Fokker-Planck equations following from
different Langevin prescriptions have been given in
Ref.~\cite{Dunkel:2009tla}. More recently, a Lorentz-covariant
implementation of the Langevin process has been discussed in
Ref. \cite{Koide:2011yy} based on a relativistic fluctuation-dissipation
theorem.

In the present work we pursue a slightly different approach. We start
from a Fokker-Planck equation with transport coefficients obtained from
an underlying microscopic theory, in the sense of an approximate 
treatment of the collision integral in the transport equation. After
recovering the generalized relativistic Einstein relation for these
coefficients, we discuss their implementation into different Langevin
realizations and verify their uniqueness in explicit examples, in
particular including the case of expanding media. We also show how
variations in the Langevin coefficients induce modifications to the
long-time limit (equilibrium) distribution of the Brownian
particle. For definiteness, our numerical applications will be
illustrated in the context of HQ diffusion in an expanding QGP as formed
in URHICs.

Our article is organized as follows. In Sec.~\ref{sec_FP} we introduce
the Fokker-Planck equation as an approximate description of the
Boltzmann equation, and recall the general equilibrium condition
constraining its drag and diffusion coefficients. In
Sec.~\ref{sec_Langevin} we elaborate how Langevin prescriptions emerge
from the pertinent equilibrium conditions to ensure a uniform outcome of
the Fokker-Planck framework, i.e., how Fokker-Planck equations resulting
from different Langevin prescriptions take the same form in terms of
universal drag and diffusion coefficients of an underlying microscopic
theory. In Sec.~\ref{sec_num} Langevin prescriptions are illustrated
with numerical calculations for different model systems for flowing
media. We demonstrate that both pre- and post-point Langevin schemes
lead to the Lorentz-invariant Boltzmann-J{\"u}ttner distribution for
test particles in the presence of collective flow as the long-time limit
of the Langevin process, if and only if the proper equilibrium condition
is imposed. Some variants of the realization of the stochastic Langevin
process and their consequences on the long-time limit are also discussed
in this section. We summarize and conclude in Sec.~\ref{sec_sum}.

\section{Fokker-Planck equation and general equilibrium condition}
\label{sec_FP}

When a heavy particle is immersed into a medium of light constituents at
small to moderate temperature, $T \lesssim m$, its momentum change due to
collisions is relatively small. Employing this soft-scattering
approximation~\cite{Landau1981}, the Boltzmann integro-differential
equation describing the motion of the heavy particles in an equilibrated
background medium reduces to the Fokker-Planck equation for the
phase-space distribution function, $f$, of the ``Brownian''
particle~\cite{Landau1981}
\begin{equation}
\label{FP}
\frac{\partial f(t,\bvec p)}{\partial t}=\frac{\partial}{\partial
p_i} \left \{ A_i(\bvec p)f(t,\bvec p)+\frac{\partial}{\partial
p_j} \left [ B_{ij}(\bvec p)f(t,\bvec p) \right] \right \} \ 
\end{equation}
with $i,j \in \{1,2,3\}$. We focus on a spatially homogeneous static
medium without external force, and thus the phase-space density is
independent of $\bvec x$.  The drag and diffusion coefficients in
Eq.~(\ref{FP}), $A_i(\bvec p)$ and $B_{ij}(\bvec p)$, respectively, are
obtained from an average of the moments of momentum transfer in
heavy-light collisions, weighted by a transition probability given by
the pertinent scattering matrix elements, $\mathcal{M}$, and the medium
particle distribution, $f^p$,
\begin{align}
\begin{split}
\label{Ai}
A_i(\bvec p) &= \frac{1}{2E(\bvec p)}\int \frac{\dd^3\bvec q}{(2\pi)^3
2E(\bvec q)}\int \frac{\dd^3\bvec q'}{(2\pi)^3 2E(\bvec q')} 
\\
& \quad \times\int \frac{\dd^3\bvec p'}{(2\pi)^3 2E(\bvec p')}
\frac{1}{\gamma}\sum\mid\mathcal{M}\mid^2 \hat{f}(\bvec q) 
\\
& \quad (2\pi)^4\delta^4(p+q-p'-q') \ [(\bvec p'-\bvec p)_i]
\\
&\equiv  \langle\langle (\bvec p'- \bvec p)_i \rangle\rangle \ , 
\end{split}
\\
\label{Bij}
B_{ij}(\bvec p)&=\frac{1}{2} \left \langle\langle (\bvec p'-\bvec p)_i(\bvec p'-\bvec
p)_j \right \rangle\rangle \ .
\end{align}
Here, $\bvec p$ and $\bvec p'$ ($\bvec q$ and $\bvec q'$) are the heavy
(light) particle's momentum before and after the collision,
respectively, and $\hat{f}=f^p(1\pm f^p)$ for bosons/fermions in the
medium.

Defining a current
\begin{equation}
\label{flux}
S_i(t,\bvec p)=-\left \{ A_i(\bvec p)f(t,\bvec p)+\frac{\partial}{\partial
p_j} \left [B_{ij}(\bvec p)f(t,\bvec p) \right] \right\} \ ,
\end{equation}
the Fokker-Planck equation can be cast as a continuity equation in
momentum space~\cite{Landau1981},
\begin{equation}\label{continuity}
  \frac{\partial f_Q(t,\bvec p)}{\partial t}+\frac{\partial}{\partial
    p_i}S_i(t,\bvec p)=0 \ .
\end{equation}
This identifies $S_i$ as a particle-number current (flux), so that the
number of heavy particles is conserved in the diffusion process. After
sufficiently many collisions with the light partons, i.e., in the
long-time limit, we expect the heavy particle to approach the same
equilibrium distribution as for the medium constituents given by the
relativistic Boltzmann-J{\"u}ttner distribution~\cite{Juttner1911},
\begin{equation}
\label{BJ}
  f_{\rm eq}(p,T)=N \exp[-E(p)/T] \ , 
\end{equation}
where $E(p)=\sqrt{\bvec p^2+m^2}$ is the relativistic on-shell energy of
the Brownian particle and $T$ the temperature of the equilibrated
medium. In addition, in statistical equilibrium, the particle flux has
to vanish~\cite{Landau1981}: $S_i(\bvec p)=0$. Together with the
Boltzmann-J{\"u}ttner distribution (\ref{BJ}) this yields a
dissipation-fluctuation relation between drag and diffusion coefficient,
\begin{equation}
\label{FPequil}
A_i(\bvec p,T)=B_{ij}(\bvec p,T)\frac{1}{T}\frac{\partial
E(p)}{\partial p_j}-\frac{\partial B_{ij}(\bvec p,T)}{\partial p_j} \ .
\end{equation}
This is the manifestation of the detailed-balance property of the
collision-transition probabilities, which is due to the unitarity of the
$S$-matrix of quantum-field theory and thus embodied in the collision
integral of the underlying full Boltzmann
equation~\cite{Landau1981}. This relation is not a priori fulfilled by
Eqs.~(\ref{Ai}) and (\ref{Bij}), and may suffer in accuracy if the
assumption of a forward-peaked transition-matrix element is not well
satisfied.  In order for the heavy particle to approach the same
distribution as for the medium particles, the drag and diffusion
coefficients are not independent but have to be related to each other
precisely as specified by the dissipation-fluctuation relation
(\ref{FPequil}). This general equilibrium condition plays a central role
in the following discussions.

\section{Stochastic Langevin Realization of Fokker-Planck Diffusion}
\label{sec_Langevin}

\subsection{Langevin Simulation}
\label{ssec_LvsFP}

The Fokker-Planck description of diffusion can be realized by Langevin's
stochastic differential
equation(s)~\cite{NGvanKampen2003,Dunkel:2009tla,Rapp:2009my}.
Employing a spatially homogeneous static medium, we follow the notation
in Ref.~\cite{Rapp:2009my} to write the Langevin equations as
\begin{alignat}{2}
\dd x_j &=\frac{p_j}{E}\dd t \ ,
\label{Lrule1}
\\
\dd p_j &=-\Gamma(p,T)p_j\dd t+\sqrt{\dd t}C_{jk}(\bvec p+\xi d\bvec
p,T)\rho_k \ ,
\label{Lrule2}
\end{alignat}
which specify the rules for updating the coordinate and momentum of the
heavy particle in time steps, $\dd t$. In Eq.~(\ref{Lrule2}),
$\Gamma(p,T)\bvec{p}$ is the deterministic friction force, whereas the
$C_{jk}$ describe the stochastically fluctuating force with independent
Gaussian noises $\rho_k$ following a normal distribution, $P(\bvec
\rho)=(2\pi)^{-3/2}e^{-\bvec\rho^2/2}$.  Thus, there is no correlation
in the stochastic forces at different times, $ \langle F_j(t)F_k(t')
\rangle=C_{jl}C_{kl}\delta(t-t')$, consistent with the assumption of
uncorrelated momentum kicks underlying the Fokker-Planck equation
(``white noise''). However, it is not specified at what momentum
argument the covariance matrix should be evaluated:
$C_{jk}=C_{jk}(t,\bvec p+\xi \dd \bvec p)$, with $\xi \in [0,1]$, e.g.,
$\xi=0$ for pre-point (Ito)~\cite{Ito1951}, $\xi=1/2$ for mid-point
(Stratonovich), or $\xi=1$ for post-point
(H{\"a}nggi-Klimontovich)~\cite{DunkeHanggi2005ab} realizations of the
stochastic integral. We will return to this point below.

The phase-space distribution determined by the Langevin equations
(\ref{Lrule1}) and (\ref{Lrule2}) satisfies a Fokker-Planck equation,
which can be found by calculating the average change of an arbitrary
phase-space function with time~\cite{Rapp:2009my}. The resulting
equation reads
\begin{equation}
\begin{split}
\label{LangevinFPtype}
\frac{\partial f(t,\bvec p)}{\partial t}=\frac{\partial}{\partial
p_j} \left [\left(\Gamma(p)p_j-\xi C_{lk}(\bvec p)\frac{\partial C_{jk}(\bvec
p)}{\partial p_l}\right) f(t,\bvec
p) \right ] \\
+\frac{1}{2}\frac{\partial^2}{\partial p_j\partial
p_k} \left[ C_{jl}(\bvec p)C_{kl}(\bvec p)f(t,\bvec p) \right].
\end{split}
\end{equation}
Comparing Eqs.~(\ref{LangevinFPtype}) and (\ref{FP}) leads to
\begin{alignat}{2}
\label{AGamma} A_j(\bvec p) &=A(p)p_j=\Gamma(p)p_j -\xi C_{lk}(\bvec
p)\frac{\partial
C_{jk}(\bvec p)}{\partial p_l},\\
\begin{split}
B_{jk}(\bvec p) &=B_0(p)P^{\bot}_{jk}(\bvec p)+B_1(p)P^{\|}_{jk}(\bvec
p) \\
&=\frac{1}{2}C_{jl}(\bvec p)C_{kl}(\bvec p) \label{BCGamma} \ , 
\end{split}
\end{alignat}
where $P^{\bot}_{jk}(\bvec p)=\delta_{jk}-p_jp_k/p^2$ and
$P^{\|}_{jk}(\bvec p)=p_jp_k/p^2$ are the corresponding projection
operators, so that
\begin{equation}
\label{CB0B1}
C_{jk}(\bvec p)=\sqrt{2B_0(p)}P^{\bot}_{jk}(\bvec
p)+\sqrt{2B_1(p)}P^{\|}_{jk}(\bvec p) \ .
\end{equation}

Two comments are in order here. First, note that the friction
coefficient, $\Gamma(p,T)$, appearing in the Langevin equation
(\ref{Lrule2}), is a derived quantity, given in terms of the drag and
diffusion coefficients, $A(p,T)$, $B_0(p,T)$ and $B_1(p,T)$, as
determined by microscopic scattering-transition matrix elements (e.g.,
they relate to heavy-quark energy loss due to elastic heavy-light
collisions~\cite{BraatenThoma1991}).  Second, the Fokker-Planck equation
and the Langevin equation are specified for a spatially homogeneous
static medium; neither of them is manifestly Lorentz covariant. Thus, in
the case of a flowing medium, momentum updates in the Langevin equation
(\ref{Lrule2}) have to be calculated in the local fluid-rest frame and
then boosted back to the moving frame when performing the diffusion
simulations. It is readily proved that the coordinate updates
(\ref{Lrule1}) can be equivalently done in the moving frame; hereafter,
we refer to the latter as the laboratory frame.

\subsection{Evaluation of Equilibrium Condition}
\label{ssec_eq-cond}

We now proceed to analyze the equilibrium condition for two
implementations of the Langevin process, the pre-point and
post-point prescriptions, and work out their manifestations in the
Fokker-Planck framework. For the sake of brevity, but without loss of
generality, we work with a diagonal approximation of the diffusion
coefficient from here on. Putting $B_0(p)=B_1(p)=D(p)$, the covariance
matrix (\ref{CB0B1}) reduces to $C_{jk}(\bvec
p)=\sqrt{2D(p)}\delta_{jk}$. Utilizing Eq.~(\ref{AGamma}), the general
equilibrium condition (\ref{FPequil}) simplifies to
\begin{equation}
A(p) = \frac{1}{E(p)}\left(\frac{D[E(p)]}{T}
        -\frac{\partial D[E(p)]}{\partial E}\right) \ 
\label{Ap}
\end{equation}
with 
\begin{equation}
\Gamma(p)= \frac{1}{E(p)}\left( \frac{D[E(p)]}{T} 
          -(1-\xi) \frac{\partial D[E(p)]}{\partial E} \right) \ .
\label{Gammap}
\end{equation}
Thus, for a given drag coefficient $A$, Eq.~(\ref{Ap}) should be solved
to obtain the diffusion coefficient $D$, which can then be used to
deduce the friction force $\Gamma$ from Eq.~(\ref{Gammap}). Alternative
procedures have been adopted in the literature by, e.g., calculating $D$
and adjusting $A$~\cite{MooreTeaney2005,Beraudo:2009pe}. 
Condition (\ref{Gammap}) has also been derived as the generalization
of the fluctuation-dissipation theorem with Boltzmann-J{\"u}ttner
distributions as the equilibrium solution in \cite{Koide:2011yy}.

In the non-relativistic limit, both $D(p)=D$ and $\Gamma(p)=\gamma$
become independent of $p$, and $E(p)\rightarrow m$; the equilibrium
condition reduces to
\begin{equation}
D=m\gamma T  \ ,
\end{equation}
which is Einstein's classical fluctuation-dissipation
relation~\cite{Landau1981}.

In terms of the diagonal diffusion coefficient, the Langevin updating
rules now read
\begin{align}
\label{Ldia-rule1}
\dd x_j &= \frac{p_j}{E}\dd t \ ,
\\
\label{Ldia-rule2}
\dd p_j &= -\Gamma(p)p_j\dd t+\sqrt{2\dd t\, D(|\bvec p+\xi d\bvec
  p|)} \rho_j \ .
\end{align}
In the following, we discuss their explicit forms in the pre-point and
post-point Langevin prescriptions as following from the corresponding
equilibrium condition, and show that they can both be rendered
compatible for a given microscopic model.

\subsubsection{Pre-point scheme: $\xi=0$}
\label{sssec_pre}

In this scheme, the equilibrium conditions (\ref{Ap}) and (\ref{Gammap})
read
\begin{equation}
\Gamma(p)= \frac{1}{E(p)}\left( \frac{D[E(p)]}{T}
          - \frac{\partial D[E(p)]}{\partial E} \right) =A(p) \ .
\label{prepoint-equil}
\end{equation}
Note that the friction force entering the Langevin equation is equal to
the drag coefficient defined by Eq.~(\ref{Ai}).

Inserting $C_{jk}(\bvec p)=\sqrt{2D(p)}\delta_{jk}$ into
Eq.~(\ref{LangevinFPtype}), the Fokker-Planck equation realized in the
pre-point Langevin scheme takes the form
\begin{equation}\label{prepointFP}
\frac{\partial}{\partial t}f(t,\bvec p)=\frac{\partial}{\partial
p_i} \left \{ \Gamma(p)p_i f(t,\bvec p)+\frac{\partial}{\partial
p_i}[D(p)f(t,\bvec p)] \right \} \ .
\end{equation}

The corresponding Langevin updating rules now read
\begin{alignat}{2}
\label{prepointLangevinrule1}
\dd x_j &= \frac{p_j}{E}\dd t \ ,
\\
\label{prepointLangevinrule2}
\dd p_j &=-\Gamma(p)p_j\dd t+\sqrt{2\dd tD(p)}\rho_j  \ ,
\end{alignat}
where carrying out the Langevin time steps is straightforward: starting
with initial coordinate $\bvec x$ and momentum $\bvec p$ of a test
particle, calculate the coordinate increment $\dd x_j$ within the time
step $\dd t$ using Eq.~(\ref{prepointLangevinrule1}); at the same time
substitute the drag coefficient $A(p)$ ($=\Gamma(p)$) evaluated at
$p=|\bvec p|$ and the corresponding $D(p)$ calculated from
Eq.~(\ref{Ap}) into Eq.~(\ref{prepointLangevinrule2}) to calculate the
momentum increment.  Then go to the next time-step with (new) initial
coordinate $\bvec x + \dd \bvec x$ and momentum $\bvec p+\dd \bvec p$.

\subsubsection{Post-point scheme: $\xi=1$}
\label{subsubpostpoint}

In this scheme, the equilibrium conditions read
\begin{align}
\label{postpointequilibirumcondition}
D[E(p)]&=\Gamma(p)E(p)T \ , \\
\label{postpoint-Gam}
\Gamma(p)&=A(p)+\frac{1}{E(p)}\frac{\partial
D[E(p)]}{\partial E} \ .
\end{align}
Here, the scheme-dependent relation takes a simple form, but the
``price'' to pay is that the friction force $\Gamma(p)$ figuring into the
Langevin equation is different from the drag coefficient $A(p)$.  In
this sense, starting from a microscopic model for $A(p)$, there is no
difference in the choice of the pre-point or the post-point scheme.

The corresponding Fokker-Planck equation in the post-point scheme takes
the form
\begin{equation}
\label{postpointFP}
\frac{\partial}{\partial t}f(t,\bvec p)=\frac{\partial}{\partial
p_i} \left \{ \Gamma(p)p_i f(t,\bvec p)+D(p)\frac{\partial}{\partial
p_i}f(t,\bvec p) \right \}\ .
\end{equation}
This Fokker-Planck equation appears to be different from its counterpart
realized in the pre-point realization, Eq.~(\ref{prepointFP}).  However,
upon substituting the different expressions for $\Gamma$,
Eqs.~(\ref{prepoint-equil}) and (\ref{postpoint-Gam}), for the pre-point
and the post-post Langevin scheme into Eqs.~(\ref{prepointFP}) and
(\ref{postpointFP}), respectively, one immediately sees that the
Fokker-Planck equations realized in the pre- and post-point Langevin
schemes are actually the same in terms of $A(p)$ and $D(p)$. This is a
desired result since the (approximate) realization of the underlying
full Boltzmann equation (including the asymptotic solution for the
distribution of test particles) within a Fokker-Planck equation should
be independent of the implementation of the stochastic process (i.e.,
Langevin scheme).  Thus, the ``Ito-Stratonovich dilemma'', which
suggests that different Langevin prescriptions are not sufficient in
order to uniquely determine the Fokker-Planck equation, does not appear
here, as a benefit of a well-defined microscopic process.

The Langevin updating rules in the post-point scheme read
\begin{alignat}{2}
\label{postpointLangevinrule1}
\dd x_j &=\frac{p_j}{E}\dd t \ ,
\\
\label{postpointLangevinrule2}
\dd p_j &=-\Gamma(p)p_j\dd t+\sqrt{2\dd tD(|\bvec p+d\bvec
p|)}\rho_j \ .
\end{alignat}
Starting with initial coordinate $\bvec x$ and momentum $\bvec p$, the
coordinate update is trivial at given time step $\dd t$, but the momentum 
update involves a two-step computation: first use the pre-point scheme 
Eq.~(\ref{prepointLangevinrule2}) to calculate the momentum increment 
$\dd \bvec p$ with $\Gamma(p)$ and $D(p)$; then evaluate the diffusion
coefficient $D$ at the argument $|\bvec p+\dd \bvec p|$ and calculate
$\dd p_j^{\text{diffusion}}\equiv\sqrt{2\dd tD(|\bvec p+d\bvec
  p|)}\rho_j$, while $\dd p_j^{\text{drag}}=-\Gamma(p)p_j\dd t$ has
already been calculated in the first step; finally, add up $\dd
p_j^{\text{diffusion}}$ and $\dd p_j^{\text{drag}}$ to obtain the total
momentum increment for the present time-step.  Note that in the two-step
momentum update procedure, the same $\dd t$ and $\rho_j$ should be used. 
A variant where different $\dd t$'s are used will be discussed in the 
case of a flowing medium in Secs.~\ref{ssec_1D-flow} and 
\ref{ssec_elliptic} below.

\section{Numerical Calculations with Different Background Media}
\label{sec_num}

In this section we perform numerical simulations of the two different
Langevin prescriptions discussed above (and some variants thereof
previously employed in the literature), to explicitly examine their
consequences for the long-time limits, in particular in the case of
flowing media. For definiteness we will adopt parameter values for
masses, temperature and flow profiles representing the problem of
heavy-quark diffusion in a QGP. We will start with a homogeneous static
medium (Sec.~\ref{ssec_static}), followed by a simple one-dimensional
flow scenario (Sec.~\ref{ssec_1D-flow}) and finally study a more
realistic elliptically expanding fireball (Sec.~\ref{ssec_elliptic}). In
all cases we use large drag and diffusion coefficients and perform the
simulations until the stationary state has been reached, demonstrating
their universal convergence to the analytical solutions for the
corresponding Fokker-Planck equations.

\subsection{Homogeneous Static Medium}
\label{ssec_static}
Let us assume a homogeneous static medium with temperature
$T=0.18\;\GeV$ with a Brownian particle of mass $m=1.5~{\rm GeV}$
immersed (we work in natural units where the velocity of light, Planck
constant, and Boltzmann constant are equal to unity:
$c=\hbar=k_{\mathrm{B}}=1$). The preceding Fokker-Planck and Langevin
equations can be readily applied; we focus on the projection on the
$z$-component of the momentum distribution, $\dd N/\dd p_z$ of the test
particle. It has been checked that the equilibrium conditions are
reached independent of arbitrarily chosen initial conditions, as
expected.

\begin{figure}[!t]
\includegraphics[width=\columnwidth]{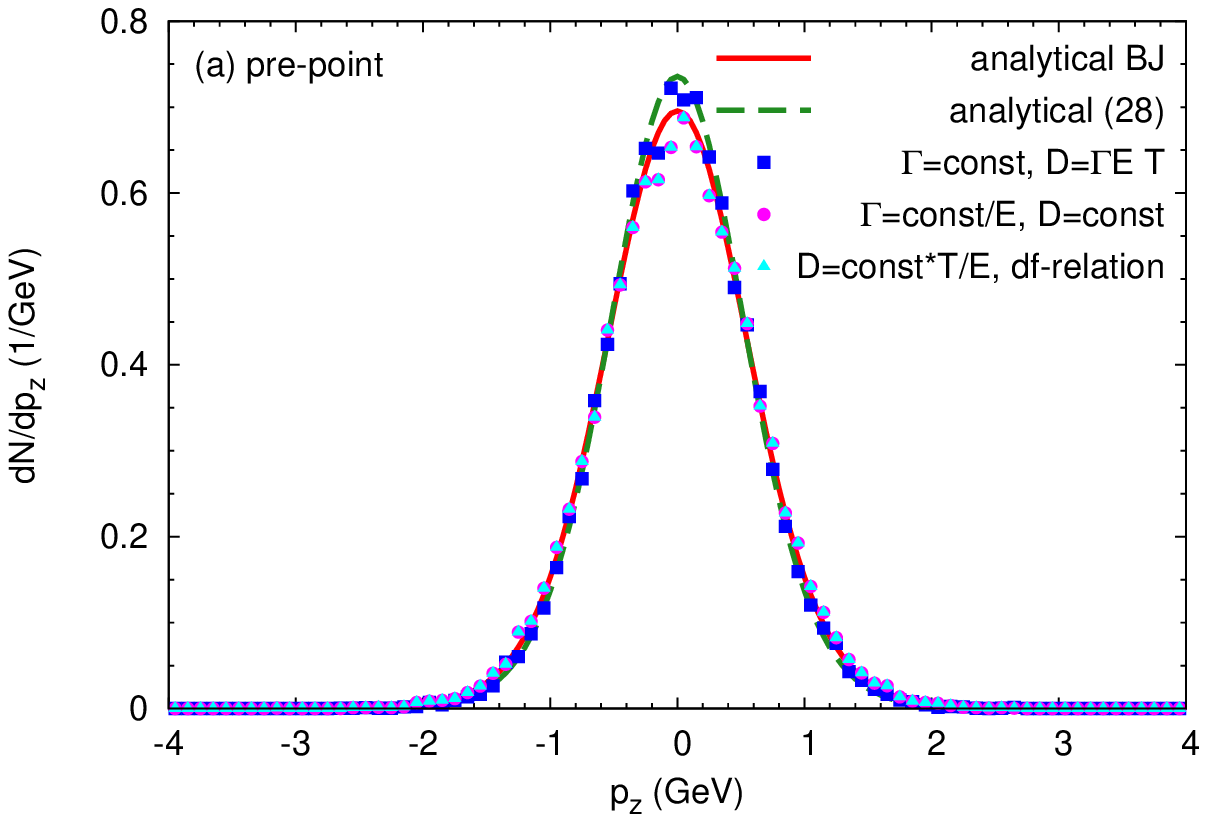} 
\includegraphics[width=\columnwidth]{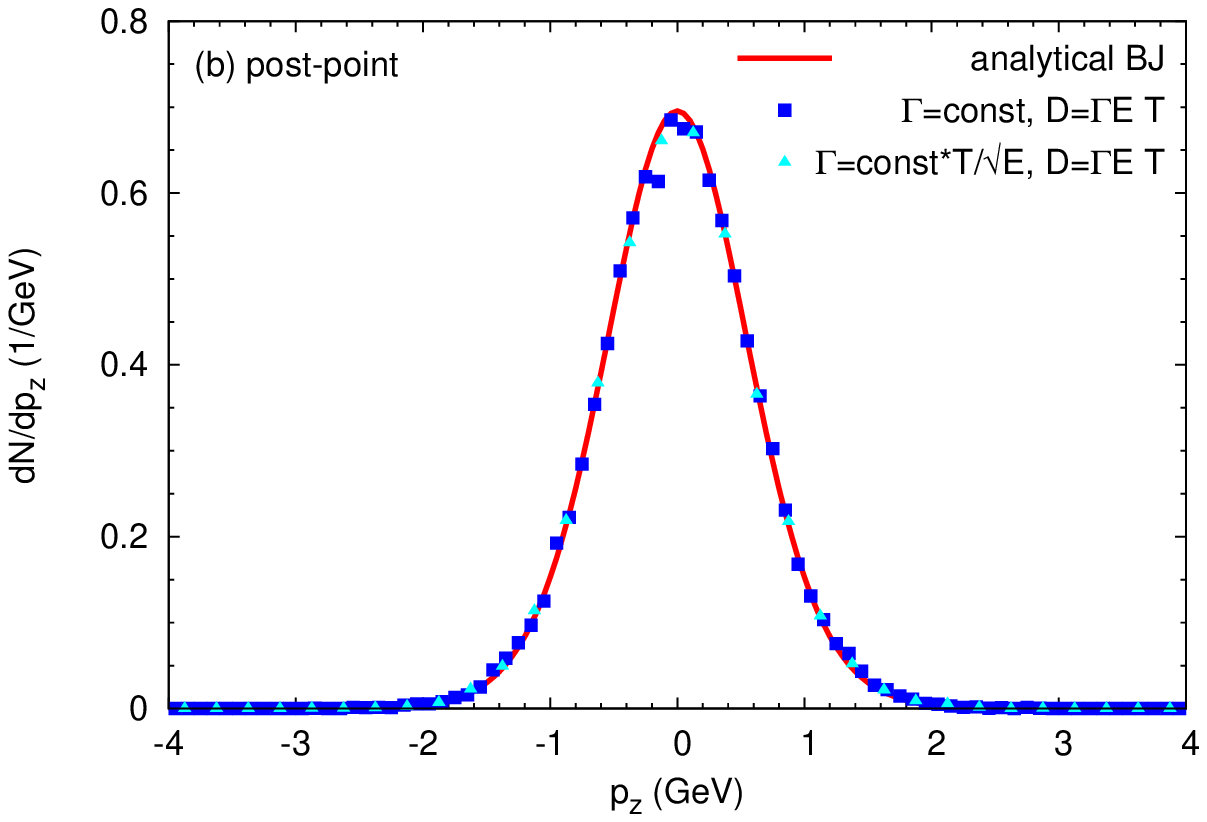} 
\caption{The distribution, $\dd N/\dd p_z$, from Langevin simulations
  for heavy quarks with mass $m=1.5 \; \GeV$, diffusing in a static
  medium at temperature $T=0.18 \; \GeV$, compared to calculations
  with the corresponding analytical phase-space distributions: (a)
  pre-point Langevin scheme, (b) post-point Langevin scheme. See the
  text for more details.} 
\label{fig_static}
\end{figure}
We first adopt the pre-point scenario in a simplistic scenario where a
large constant friction coefficient ($\Gamma=20\,\fm^{-1}$) is
complemented by a diffusion coefficient $D=\Gamma E T$.  This leads to a
stationary distribution,
\begin{equation}
  \dd N/\dd^3x\dd^3p=m/E \exp[-E(p)/T], 
\end{equation}
as shown in the upper panel of Fig.~\ref{fig_static}. It can be easily
verified analytically by plugging the drag and diffusion coefficients
into the Fokker-Planck equation (\ref{prepointFP}) and performing the
momentum derivatives, as was also discussed in
Ref.~\cite{DunkeHanggi2005ab}. However, this form of drag and diffusion
coefficient does \textit{not} comply with the pre-point equilibrium
condition (\ref{prepoint-equil}), and thus the simulation fails to
converge to the Boltzmann-J{\"u}ttner distribution.

To illustrate the importance of implementing the proper equilibrium
condition, we have performed Langevin simulations within the pre-point
scheme
using two types of momentum dependencies of the diffusion coefficient:
\begin{align}
\text{(i) } & \Gamma=20/E \; \GeV/\fm \, ; \label{eq:coeff1}\\ 
               & D=\Gamma E T=20 \, T\; \GeV/\fm \, .  \nonumber \\
\text{(ii) } & \Gamma=300(1+T/E)/E^2 \; \GeV^2/\fm  \, ;  \label{eq:coeff2}\\
  & D=300 \, T/E \; \GeV^2/\fm \, . \nonumber
\end{align}
The lifetime of the system in these simulations has been set to $10 \; \fm$,
and it has been checked that the stationary distribution has
been reached.

The simulated results are also shown in the upper panel of
Fig.~\ref{fig_static}. Again, their agreement with the
Boltzmann-J{\"u}ttner distribution can be analytically verified by
substituting the coefficients into the Fokker-Planck equation
(\ref{prepointFP}) and carrying out the momentum derivatives.

Next, we employ the post-point scenario. The simulated results are shown
in the lower panel of Fig.~\ref{fig_static}. Here, the Langevin
simulations with large drag reproduce well the Boltzmann-J{\"u}ttner
distribution, whether the friction is constant ($\Gamma=20/\fm$,
$D=\Gamma E T$) or not ($\Gamma=40 \sqrt{\GeV}\,\fm^{-1}/\sqrt{E(p)}$,
$D=\Gamma E T$), since the post-point equilibrium condition
(\ref{postpointequilibirumcondition}) is always fulfilled. This is
easily verified by plugging the coefficients into the corresponding
Fokker-Planck equation (\ref{postpointFP}).

To summarize this part, we have confirmed through numerical simulations
that in order for the stationary Langevin limit to converge to the
Boltzmann-J{\"u}ttner distribution, the equilibrium condition must be
\emph{exactly} fulfilled.

\subsection{Constant 1-D Flow}
\label{ssec_1D-flow}

We now introduce a constant one-dimensional medium flow with velocity
field $v_x=0$, $v_y=0$ and $v_z=0.9$. The medium temperature,
$T=0.18\;\GeV$, and heavy-quark mass, $m=1.5\;\GeV$, are as before.

\begin{figure}[!t]
\label{plots1Dflow}
\includegraphics[width=\columnwidth
]{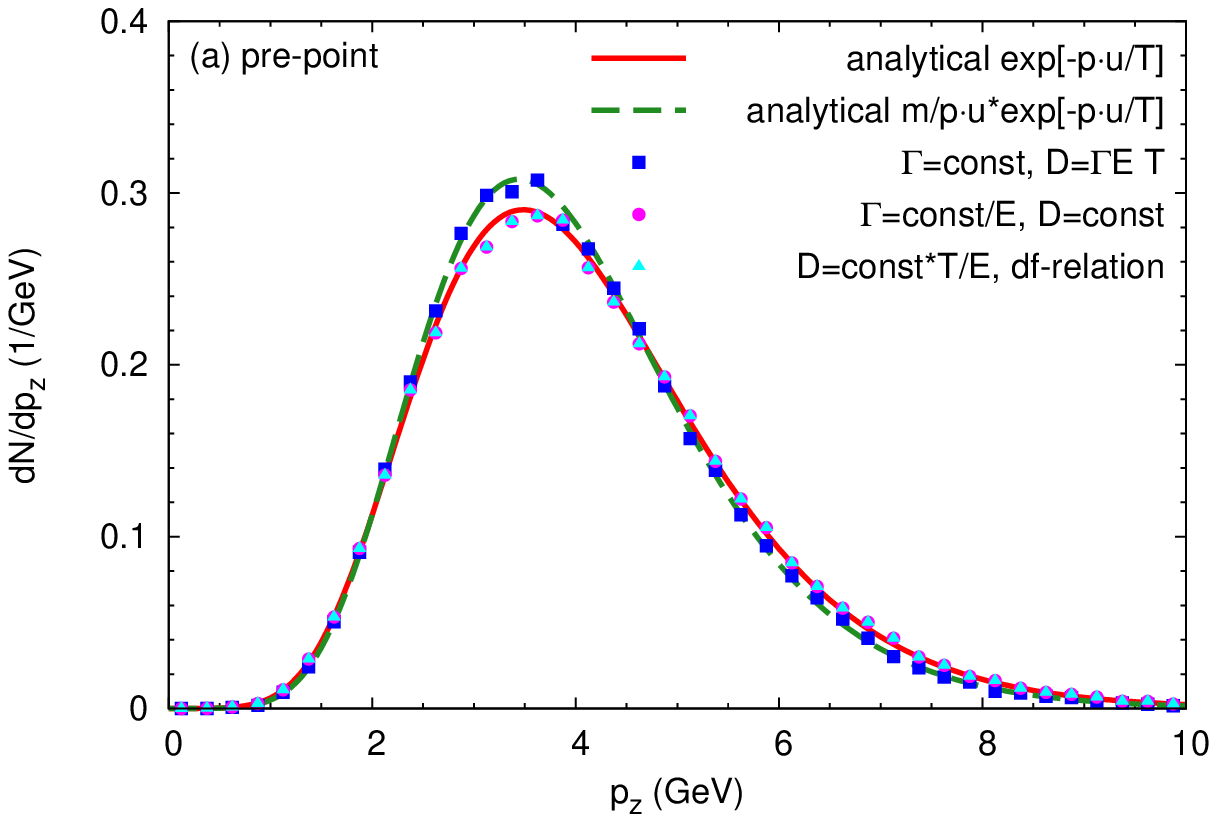}
\includegraphics[width=\columnwidth
]{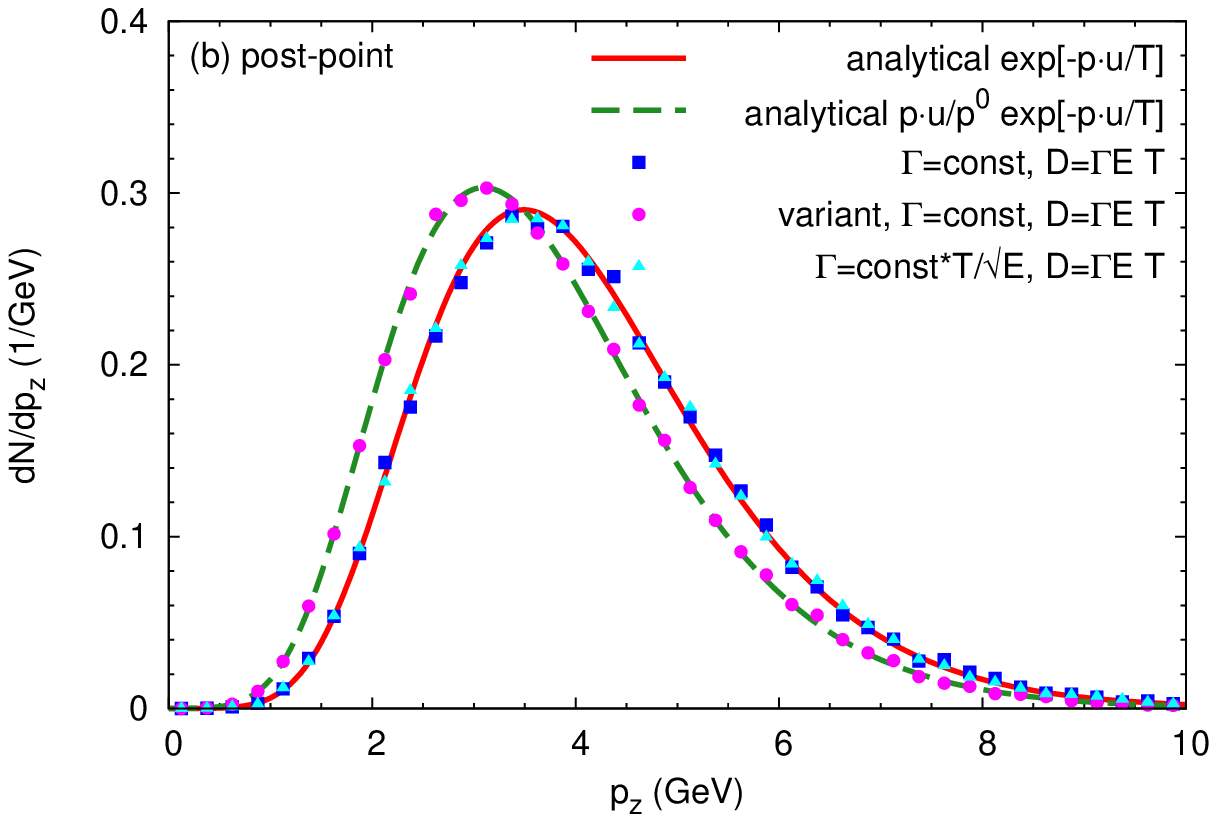} 
\caption{Langevin simulation results for heavy quarks ($m=1.5\;\GeV$)
  diffusing in a flowing medium ($T=0.18\;\GeV$, $v_z=0.9$) compared to
  calculations with analytical phase-space distributions: (a) pre-point
  Langevin scheme, (b) post-point Langevin scheme. The distribution
  obtained with a variant of post-point scheme and the corresponding
  blast-wave distribution are also shown. See text for more details.  }
\end{figure}
The numerical simulations in the pre-point scheme directly translate the
result for the static medium into the corresponding ``blast-wave''
distribution, obtained by replacing $E$ with $p\cdot u$. Specifically, a
momentum-independent drag complemented by the naive $D=\Gamma E T$ leads
to a single-particle phase-space distribution $\dd
N/\dd^3x\dd^3p=m/(p\cdot u) \exp(-p\cdot u/T)$, while general
momentum-dependent coefficients that fulfill the pre-point equilibrium
condition (\ref{prepoint-equil}) give the  Boltzmann-J{\"u}ttner
distribution, $\dd N/\dd^3x\,\dd^3p=\exp(-p\cdot u/T)$, see upper panel
of Fig.~\ref{plots1Dflow}.

For the post-point scenario, where the equilibrium condition takes the
simple form (\ref{postpointequilibirumcondition}), the momentum update
involves two steps in the fluid-rest frame. There seems to be an
ambiguity in the time increment $\dd t^*$ used in the two-step
computation (here and in the following, variables with (without)
superscript $*$ refer to the fluid rest (laboratory) frame). For the
first step, it is clear that $\dd t^*_{(1)}=\dd t\frac{E^*}{E}=\dd
t\frac{p\cdot u}{p^0}$ with $E=p^0$ being the lab-frame energy prior to
the momentum update and $E^*=p\cdot u$ the corresponding energy measured
in the fluid rest frame. One then updates the momentum and obtains
$p_{(1)}^{*\mu}=(E^*_{(1)},\bvec p^*_{(1)})$ in the fluid rest frame;
the corresponding lab-frame four-momentum, $p^{\mu}_{(1)}=(E_{(1)},\bvec
p_{(1)})$, follows from a boost using the fluid velocity $u^{\mu}$. In
the subsequent second-step momentum update, one finds a particle
phase-space distribution, $\dd N/\dd^3x\dd^3p=(p\cdot u/p^0)
\exp(-p\cdot u/T)$ in the lab frame if one uses $\dd t^*_{(2)}=\dd
t\frac{E^*_{(1)}}{E_{(1)}}=\dd t \frac{p_{(1)}\cdot u}{E_{(1)}}\neq \dd
t^*_{(1)}$. This is different from the Boltzmann-J{\"u}ttner
distribution (we denote it as a ``variant''), as illustrated in the
lower panel of Fig.~\ref{plots1Dflow}. One finds both analytically and
numerically that the Boltzmann-J{\"u}ttner distribution is recovered
when using $\dd t^*_{(2)}=\dd t^*_{(1)}$ in the second-step momentum
update.  Recalling that in the general derivation of Fokker-Planck
equation from the Langevin equations~\cite{Rapp:2009my} (for a static
thermal medium) a fixed $\dd t^*=\dd t^*_{(1)}$ is adopted, we conclude
that in the post-point scheme, one should use the same $\dd t^*=\dd
t^*_{(1)}$ for both momentum-update steps in order to obtain the
Lorentz-invariant Boltzmann-J{\"u}ttner distribution in the presence of
flow.

\subsection{Elliptic Fireball}
\label{ssec_elliptic}
\begin{figure}[!t]
\vspace{-0.6cm}
\includegraphics[width=1.1\columnwidth]{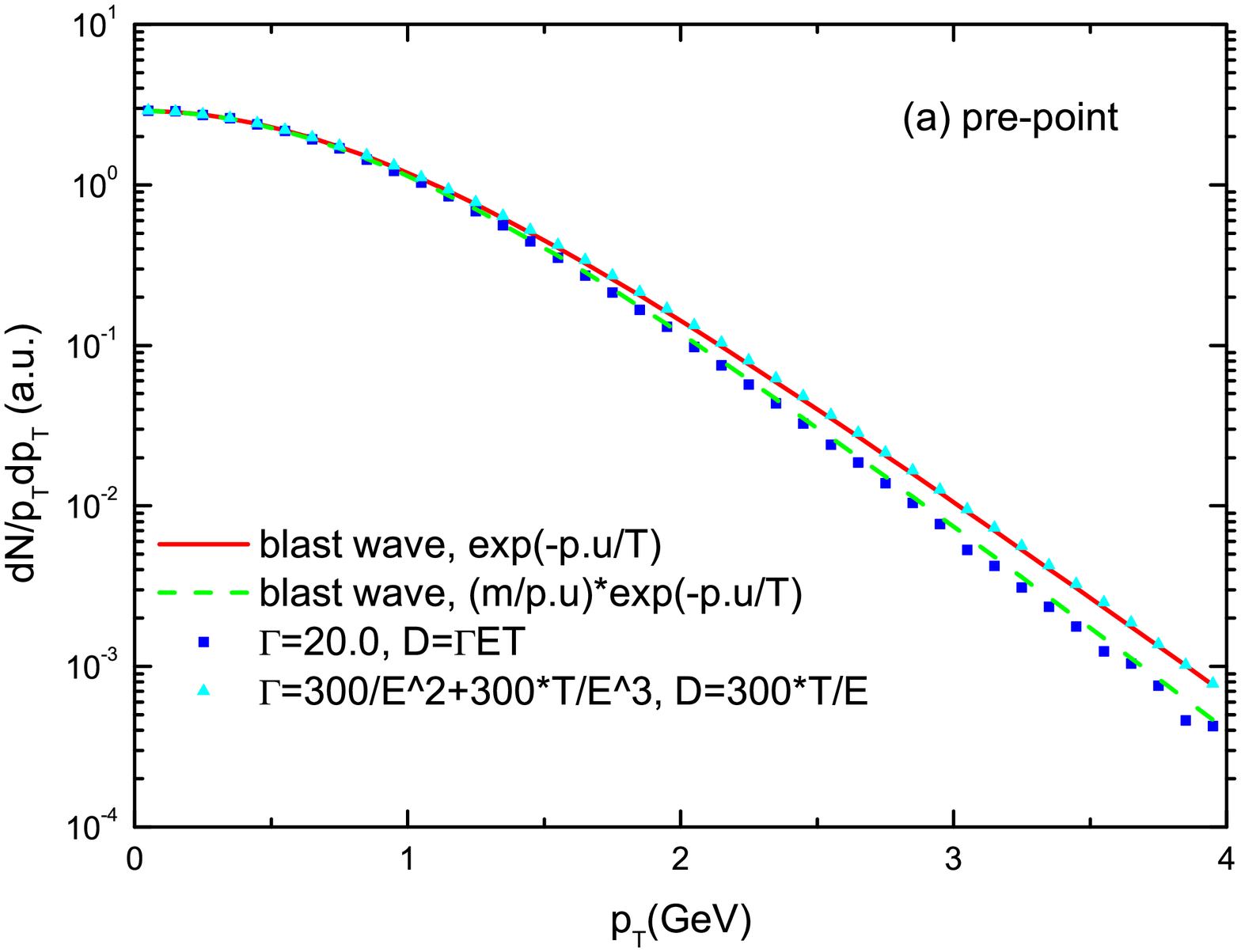}
\vspace{-0.8cm}
\includegraphics[width=1.1\columnwidth]{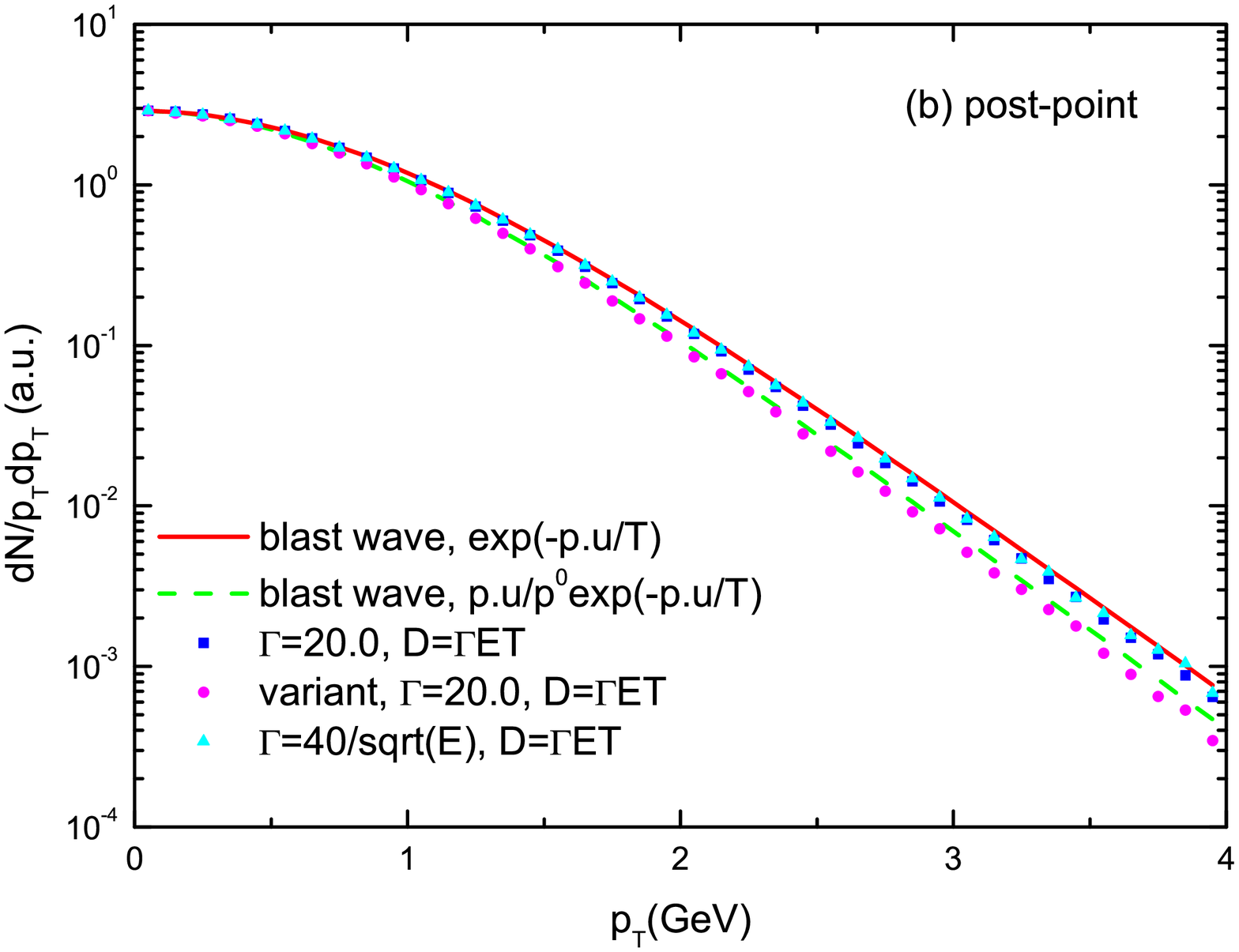}
\caption{Langevin simulation of the $p_T$-spectrum of a heavy quark
  ($m=1.5\;\GeV$) diffusing in a fireball ($T_i=0.33\; \GeV$,
  $T_f=0.18\;\GeV$), compared to the direct blast wave
  calculations. Upper panel: pre-point Langevin scheme; lower panel:
  post-point Langevin scheme. See text for more details.}
\label{fbpTspectrum}
\end{figure}
In non-central heavy-ion collisions, interactions among particles
convert the initial spatial asymmetry into particle momentum
anisotropies, most notably an ``elliptic flow'' quantified by the
coefficient of the second harmonic, $v_2(p_T)$, in the azimuthal-angle
distribution of the particle
spectrum~\cite{Ollitrault:1992bk,Shuryak2009},
\begin{equation}
\label{v2}
\frac{\dd N}{p_T\dd p_T \dd \phi_p}=\frac{\dd N}{2\pi p_T\dd p_T} [1+2v_2(p_T)
  \cos (2\phi_p)+\cdots],
\end{equation}
where $p_T$ is the transverse momentum of the emitted particle. Heavy
quarks (or hadrons containing heavy quarks) are believed to acquire a
$v_2$ through the coupling to the collective motion of the light
particles in the QGP~\cite{MooreTeaney2005,Rapp:2009my}. In the
following, we employ an elliptically expanding fireball, introduced in
Ref.~\cite{vanHees:2005wb} (see also the discussion in
Ref.~\cite{Gossiaux2011} for more details) to model the QGP medium QGP
evolution, in order to scrutinize the Langevin simulation of heavy-quark
(HQ) diffusion in the QGP background. We again take the HQ mass to be
$m=1.5\,\GeV$ and employ the same two sets of coefficients $\Gamma$ and
$D$ as specified in Eqs.~(\ref{eq:coeff1}) and (\ref{eq:coeff2}),
satisfying the equilibrium conditions. This time the Langevin simulation
runs in parallel to an isentropically expanding fireball which stops at
the decoupling temperature $T_f=0.18\,\GeV$. We compute the spectrum and
elliptic flow and compare to the results of a direct blast-wave
calculation which corresponds to the equilibrium limit of the fireball.

Let us first discuss the HQ $p_T$-spectrum, shown in
Fig.~\ref{fbpTspectrum}.  For the pre-point scenario (upper panel),
Langevin simulation results translate again directly into the
corresponding blast-wave distributions (obtained by replacing $E$ with
$p\cdot u$). For a momentum-independent drag coefficient complemented by
the naive $D=\Gamma E T$ (at variance with the equilibrium condition),
the HQ phase-space distribution assumes the form $\dd
N/\dd^3x\dd^3p=m/(p\cdot u) \exp(-p\cdot u/T)$, while general momentum
dependent coefficients that fulfill the pre-point equilibrium condition
(\ref{prepoint-equil}) accurately recover the Boltzmann-J\"uttner
distribution $\dd N/\dd^3x\dd^3p=\exp(-p\cdot u/T)$. For the post-point
scenario (lower panel), again, once the corresponding equilibrium
condition (\ref{postpointequilibirumcondition}) is satisfied, the
Langevin simulation also yields the correct Lorentz-invariant Boltzmann
distribution, while the ``variant'' scheme leads to an explicitly frame
dependent distribution $\dd N/\dd^3x\dd^3p=(p\cdot u/p^0) \exp(-p\cdot
u/T)$. This results from using a different $\dd t^*_{(2)}\neq \dd
t^*_{(1)}$ in the two-step momentum update, as discussed in
Sec.~\ref{ssec_1D-flow}.

\begin{figure}[!t]
\vspace{-0.6cm}
\includegraphics[width=1.1\columnwidth]{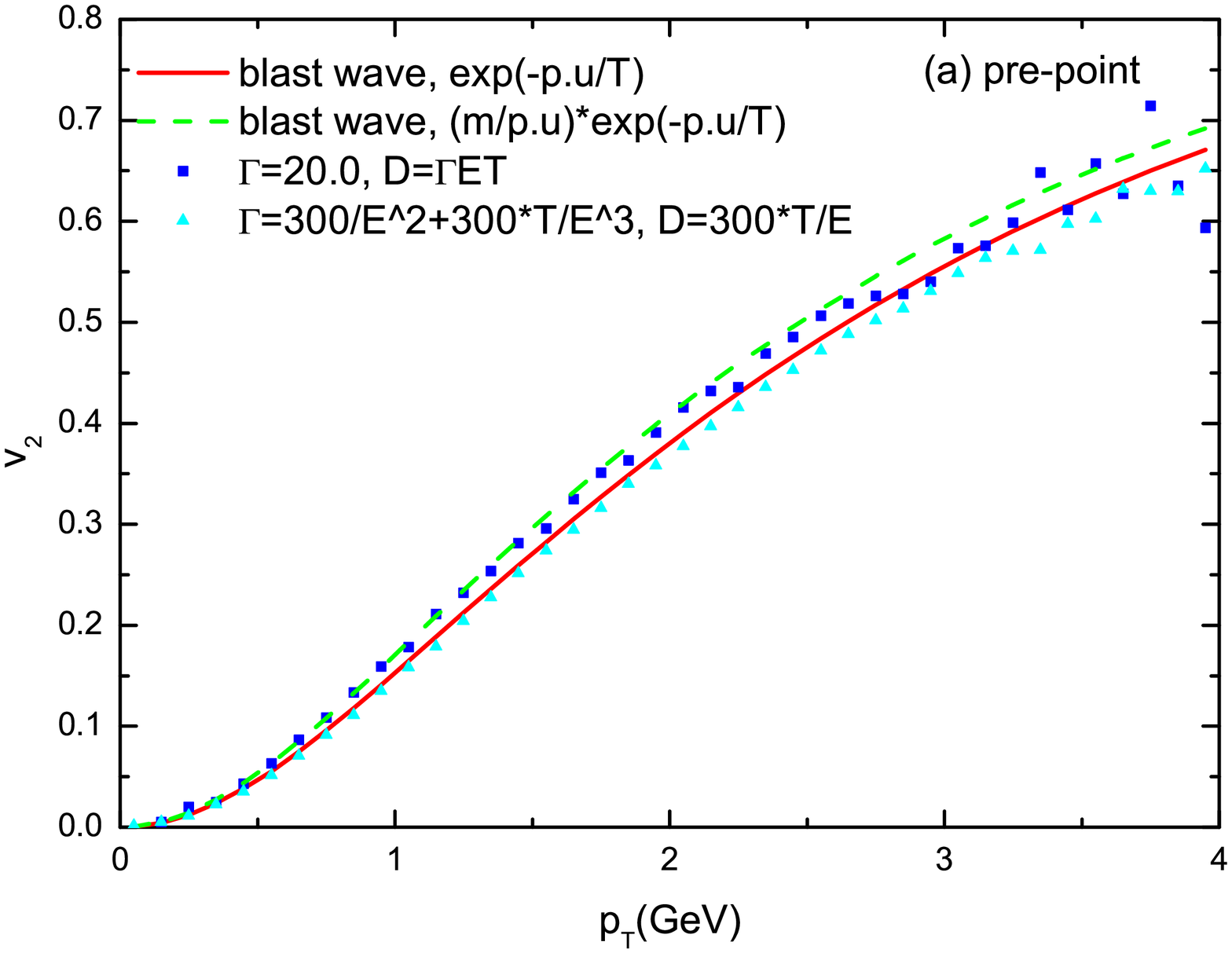} 
\vspace{-0.8cm}
\includegraphics[width=1.1\columnwidth]{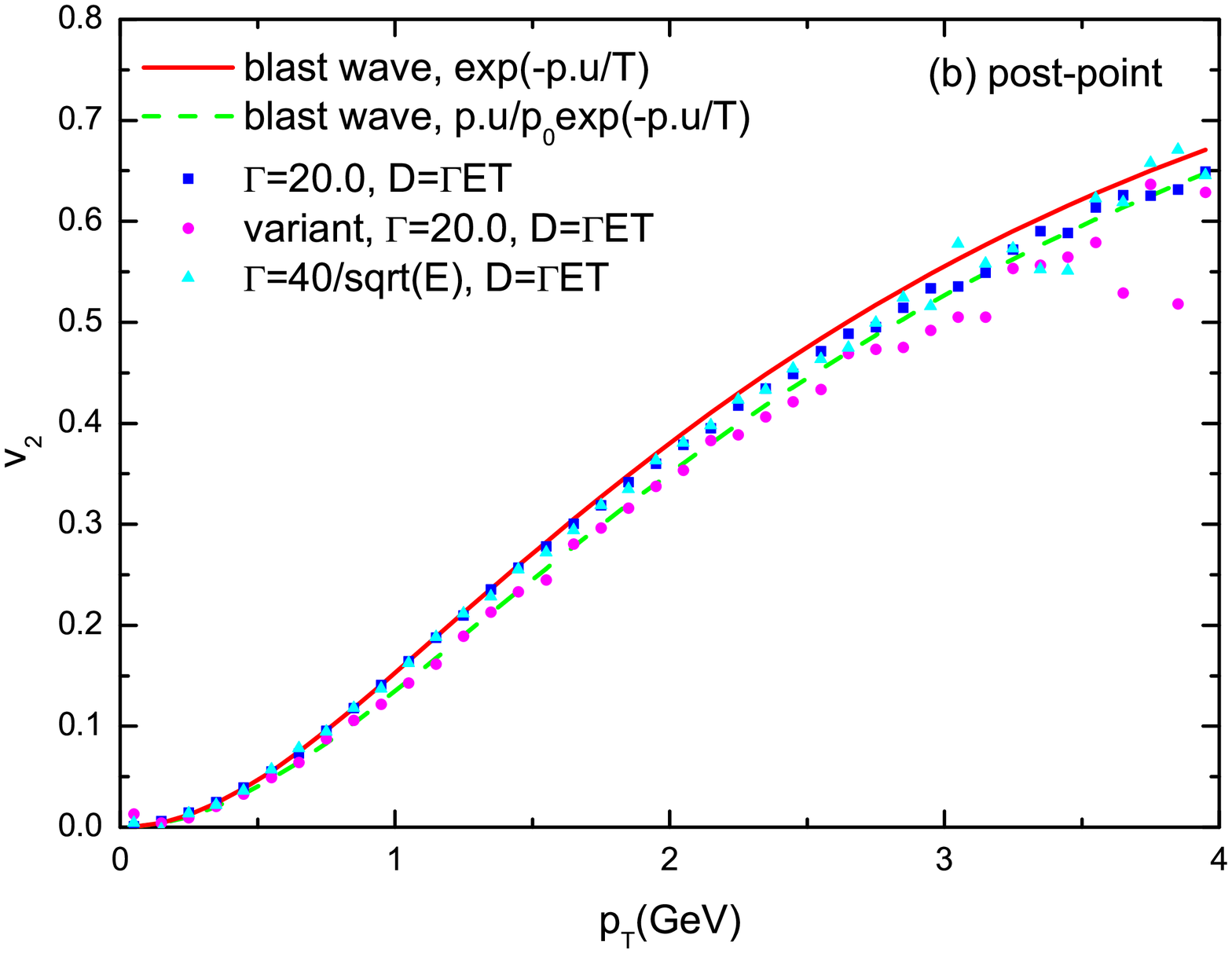} 
\caption{Langevin simulation results for $v_2(p_T)$ for a heavy quark
  ($m=1.5\;\GeV$) diffusing in a QGP ($T_f=0.18\;\GeV$) compared to
  direct blast-wave calculations with the flow field and temperature of
  the background medium (fireball). Upper panel: pre-point Langevin
  scheme; lower panel: post-point Langevin scheme.}
\label{fbv2}
\end{figure}
The HQ $v_2(p_T)$ obtained from the equilibrium limit of the Langevin
simulations and its comparison with the direct blast-wave calculations
are shown in Fig.~\ref{fbv2}. The agreement between different
implementations (pre-point, post-point and variants) of Langevin
simulations and the corresponding blast-wave distributions confirm the
conclusions for the $p_T$ spectra, reproducing accurately even rather
subtle angular modulations in relativistic flow fields.

\section{Summary and Conclusions}
\label{sec_sum}

In this work we have explored aspects of relativistic Langevin dynamics,
in particular their uniqueness relative to a Fokker-Planck description
and their manifestation in the presence of non-trivial medium-flow
fields.  Our perspective on these issues commenced from a microscopic
theory for the interactions of a Brownian particle in a fluid of light
particles, where the Fokker-Planck equation emerges from the Boltzmann
equation with well-defined transport coefficients. We thus started by
re-establishing the general constraint between the drag and diffusion
coefficients in order for the asymptotic solution of the Fokker-Planck
equation to converge to the Boltzmann-J{\"u}ttner distribution in the
equilibrium limit. Based on this ``master equation'' we investigated two
widely used Langevin realizations of the Fokker-Planck equation and
explicitly obtained the equilibrium conditions for the coefficients in
these schemes. It followed that both pre-point and post-point Langevin
equations obey an equivalent Fokker-Planck equation in terms of the
original drag and diffusion coefficients, thus illuminating the
``Ito-Stratonovich dilemma'', i.e., the pre-point and the post-point
Langevin algorithm turn out to be equally adequate to describe the same
micro-physics. We verified these results by explicit numerical
simulations, recovering the Boltzmann-J{\"u}ttner distribution as the
equilibrium limit of heavy test particles, if and only if the
corresponding equilibrium condition is implemented exactly. We
furthermore confirmed that the Langevin dynamics preserves the Lorentz
invariance of the particles' phase-space distribution in the presence of
collective-flow fields of the background medium: the particle's energy
in the fluid rest frame, $E^*$, simply converts into $p\cdot u$, where
$p$ is the particle's four momentum in the lab frame, and $u$ is the
fluid four-velocity. This, in particular, showed the fulfillment of
Lorentz-covariance through the two levels of approximation from full
Boltzmann transport to the Langevin process. We believe that these
insights are, in particular, useful for reliable simulations of
heavy-quark diffusion in heavy-ion collisions, where the coupling of the
Brownian particle to the rather subtle relativistic flow fields has
become a quantitative tool to extract transport properties of the
background medium.

\vspace{0.3cm}

\acknowledgments 
This work has been supported by U.S. National Science Foundation under 
grants CAREER PHY-0847538, PHY-0969394 and PHY-1306359, and 
by the A.-v.-Humboldt Foundation.

\end{document}